\documentstyle[12pt]{article}
\def\laq{\raise 0.4 ex \hbox{$<$}\kern -0.8 em\lower 0.62 ex\hbox{$\sim$}}
\def\gaq{\raise 0.4 ex \hbox{$>$}\kern -0.7 em\lower 0.62 ex\hbox{$\sim$}}
 
\def\beq{\begin{equation}}
\def\eeq{\end{equation}}
\def\bea{\begin{eqnarray}}
\def\eea{\end{eqnarray}}
\def\bq{\begin{quote}}
\def\eq{\end{quote}}

\def\ga{\gamma}

\def\De{\Delta}

\def\prt{\partial}

\def\half{{\textstyle{1\over 2}}}
\def\frac#1#2{{\textstyle{{#1}\over {#2}}}}

\def\lsim{\mathrel{\rlap{\lower4pt\hbox{\hskip1pt$\sim$}}
    \raise1pt\hbox{$<$}}}
\def\gsim{\mathrel{\rlap{\lower4pt\hbox{\hskip1pt$\sim$}}
    \raise1pt\hbox{$>$}}}
\def\sqr#1#2{{\vcenter{\vbox{\hrule height.#2pt
         \hbox{\vrule width.#2pt height#1pt \kern#1pt
         \vrule width.#2pt}
         \hrule height.#2pt}}}}

\def\rl{\stackrel{\leftrightarrow}{\hskip1pt\prt^{\nu}}}

\def\AJ{{\it Ap. J.} }

\def\ARAA{{\it Ann. Rev. Astron. Astrophys.} }
\def\AP{{\it Ann. Phys.} }

\def\ASAS{{\it Astron. and Astrophys.} }

\def\CQG{{\it Class. Quantum Gravity} }

\def\GRG{{\it Gen. Relativity and Gravitation} }

\def\JP{{\it J. Phys.} }

\def\PL{{\it Phys. Lett.} }
\def\PR{{\it Phys. Rev.} }
\def\PRL{{\it Phys. Rev. Lett.} }

\def\PTP{{\it Progr. Theor. Phys.} }

\def\RMP{{\it Rev. Mod. Phys.} }

\def\gappeq{\mathrel{\rlap {\raise.5ex\hbox{$>$}}
{\lower.5ex\hbox{$\sim$}}}}
 
\def\lappeq{\mathrel{\rlap{\raise.5ex\hbox{$<$}}
{\lower.5ex\hbox{$\sim$}}}}

\begin{document}

\begin{flushright}
{DF/IST - 6.2000} \\
{March 2001} \\
{\tt astro-ph/yymmnn}
\end{flushright}
\vglue 1cm
	
\begin{center}
{{\bf  Ultra-High Energy Cosmic Rays and Symmetries of Spacetime}
\footnote{This essay 
was selected for an Honorable Mention by the Gravity Research Foundation, 
2001.} \\}

\vglue 0.5cm
{O.\ Bertolami~}\footnote{E-mail address: {\tt orfeu@cosmos.ist.utl.pt}}

\bigskip
{\it Instituto Superior T\'ecnico,
Departamento de F\'\i sica,\\}
\smallskip
{\it Av.\ Rovisco Pais, 1049-001 Lisboa, Portugal\\}
\end{center}

\setlength{\baselineskip}{0.7cm}

\vglue 1cm

\centerline{\bf  Abstract}
\vglue 0.5cm
\noindent
High energy cosmic rays allow probing phenomena that 
are inacessible to accelerators. Observation of cosmic rays, presumebly 
protons,  with energies 
beyond $4 \times 10^{19}~eV$, the so-called Greisen-Zatsepin-Kuzmin (GZK) 
cut-off, give origin to two puzzles: How do 
particles accelerate to such energies ? Are their sources 
within $50 - 100~Mpc$ from Earth, or Lorentz invariance is actually 
a broken symmetry ? We suggest an astrophysical test to verify the latter 
alternative and explore a possible connection with an alternative 
theory of gravity that exhibits preferred-frame effects.

\vglue 1cm

\pagestyle{plain}

\setcounter{equation}{0}
\setlength{\baselineskip}{0.7cm}

The recent observation \cite{Hayashida,Bird,Lawrence,Efimov}
of cosmic rays with energies 
beyond the GZK cut-off 
\cite{Greisen} poses, besides the 
puzzle of understanding the mechanism that allows accelerating cosmic 
particles, presumebly protons, 
to such energies, the interesting side riddle of either locating
viable sources at distances within, $D_{Source} \lsim 
50 - 100~Mpc$ \cite{Hill}, or verifying, in an independent way, 
the violation of Lorentz invariance \cite{Sato1,Coleman,Mestres,Bertolami1}. 
Indeed, as is already well known, the latter comes about as only through the 
violation of that spacetime symmetry one can have energy-dependent effects 
that suppress processes such as 
the resonant scattering reaction with photons of the Cosmic Microwave 
Background (CMB), e.g. 
$p + \ga_{2.73K} \to \De_{1232}$, which are central in establishing 
the GZK cut-off. The astrophysical alternative is, as mentioned above, 
to identify a set of 
nearby sources so that the travelling time of the emitted particles 
is shorter than the attenuation 
time due to particle photoproduction on the CMB. Given the energy of the 
observed ultra-high energy cosmic rays (UHECRs) and the Hilla's criteria 
\cite{Hillas} on the 
energy, size and intensity of the magnetic field involved, 
$E_{18} \le {1 \over 2} \beta Z B(\mu G) L(kpc)$ - where $E_{18}$ is 
the maximum energy measured in units of $10^{18}~eV$, $\beta$ is the velocity 
of the shock wave relative to $c$ and $Z$ is the atomic number - it
implies that, within a volume of radius 
$50 - 100~Mpc$ about the Earth, only neutron stars, active galactic 
nuclei (AGN), gamma-ray bursts (GRBs) and cluster of galaxies 
are feasible acceleration sites \cite{Hillas,Cronin}. 
This astrophysical alternative has been recently advocated 
in Ref. \cite{Farrar}, where it is further argued that the near isotropy 
of the arrival directions of the observed UHECRs can be attributed to 
extragalactical magnetic fields near the Milky Way that are strong enough to 
deflect and isotropise the incoming directions of UHECRs from sources within 
$D_{Source}$ \footnote{Isotropy is a sensitive issue from the observational 
point of view \cite{Tanco} and rather crucial in the discussion about 
the origin of the UHECRs 
as scenarios for their production naturally generate anisotropies (see e.g. 
\cite{Olinto}).}. However, a serious objection against this proposal 
is the mismatch in the energy fluxes of observed UHECRs 
and of the potential sources. Indeed, assuming, for instance, 
that the energy output in UHECRs owe its origin to GRBs and is 
about $E \sim 10^{52}~erg$, a 
$\Lambda$CDM model and that the GRBs rate is proportional to the star 
formation, it follows that the resulting flux is, 
$\Phi \le 3 \times 10^{-3}~eV~cm^{-2}~s^{-1}~sr^{-1}$  \cite{Dar1}, 
about four orders of magnitude 
smaller than the observed flux of UHECRs. Similar arguments and knowledge 
of the AGN evolution function is sufficient to also exclude those objects.

Another astrophysical proposal suggests that relativistic jets from 
hypothetical GRBs in the Milk Way that do not point in our direction, 
accelerate UHECRs in our galactic halo \cite{Dar2}. No other evidence 
of this nearby population of GRBs is known and, from the knowledge 
of GRBs acquired from the recent observation of their 
afterglows, it has became clear that
GRBs all have an identifiable host galaxy. Hence, from the flux argument 
discussed above, it follows that invoking a local population of GRBs
is almost as claiming a special status for our galaxy.   

A further difficulty related with the astrophysical route is that studies 
where correlations between observed UHECRs and potential candidates were 
sought are, so far, inconclusive in what concerns correlations with 
large scale structure \cite{Waxman} and high redshift sources \cite{Sigl}. 
Of course, these negative results might be attributed to the small size of the 
available samples.

It seems, therefore, fair to conjecture that, given the abovementioned 
arguments, the breaking of Lorentz invariance should also be 
seriously considered.  
Lorentz invariance is one of the most fundamental symmetries of physics 
and is an underlying ingredient of 
all known physical descriptions of nature. However, more 
recently, there has been evidence in the context of string/M-theory 
that this symmetry can be spontaneously broken due to
non-trivial solutions in the field theory of open strings \cite{Kostelecky1} 
and interactions that may emerge in a scenario where our world 
is wrapped in a tilting $3$-brane \cite{Dvali}. The resulting novel
interactions may have striking implications at low-energy 
\cite{Bertolami2,Bertolami3}.
Putative violations of the Lorentz invariance may also lead to
the breaking of CPT symmetry \cite{Kostelecky2}. 
Interestingly, this last 
possibility can be verified experimentally in 
neutral-meson experiments \cite{Colladay1}, Penning-trap 
measurements and hydrogen-antihydrogen spectroscopy 
(see \cite{Bluhm} for an updated review). 
The breaking of CPT symmetry also allows for 
an explanation of the baryon asymmetry of the Universe \cite{Bertolami4}.
An extension of the Standard Model (SM) that incorporates possible violations 
of Lorentz and CPT 
symmetries was built in Ref. \cite{Colladay2}.

The violation of these fundamental symmetries naturally raise the question of 
how to phenomenologically test them. Astrophysics may play an 
essential role in this respect as it will soon 
be possible to make correlated
astrophysical observations involving high-energy radiation and, for instance, 
neutrinos which will make viable direct astrophysical tests of 
Lorentz invariance as we discuss next (see also \cite{Bertolami1,Amelino}) 
and references therein). 

From the experimental point of view, the most stringent
limit on the violation of Lorentz symmetry arises from measurements of 
the time dependence of the quadrupole splitting of nuclear Zeeman levels
along Earth's orbit giving origin to the impressive 
limit on deviations from the Lorentz invariance, 
$\delta < 3 \times 10^{-22}$ \cite{Lamoreaux}, 
and even more stringent bounds according to a recent assessment 
\cite{Kostelecky3}.

Bounds on the violation of Lorentz symmetry can be also
extracted from ultra-high energy cosmic rays, from which a 
limit on the difference of the maximum propagation velocity of different 
particles is obtained, e.g. for the $\Delta_{1232}$ resonance, 
$c_p - c_{\Delta} \equiv \epsilon_{p \Delta} 
\simeq 1.7 \times 10^{-25}$ \cite{Coleman} 
and from the search of neutrino oscillations, 
$\vert \epsilon \vert \lsim few \times 10^{-22}$ \cite{Brucker}.
These limits can be turned into bounds on parameters of the 
Lorentz-violating extension of the SM \cite{Bertolami1}. Actually, in the 
context of that extension, and also in some quantum gravity and  
stringy induced models (see \cite{Amelino,Mavromatos}), 
deformations on the relativistic 
dispersion relation are obtained that, give origin, for certain instances,
to threshold effects that allow evading the GZK cut-off 
\cite{Coleman,Bertolami1,Aloisio,Sato2}, to a time delay in the arrival 
of signals from faraway sources carried by different particles
\cite{Bertolami1}, and to bounds on the quantum gravity scale 
\cite{Amelino,Mavromatos}.
As discussed in \cite{Bertolami1}, a distinct feature of the Lorentz violating 
extension of the SM of Ref. \cite{Colladay2} is that it leads 
to a time delay, $\Delta t$, in the arrival of signals 
carried by different particles that is energy {\it independent}, in 
opposition to what was expected from general arguments (see \cite{Amelino} and 
references therein), and 
has a dependence on the {\it chirality} of the particles involved:

\beq
\Delta t \simeq {D \over c} [(c_{00} \pm d_{00})_{i} - 
(c_{00} \pm d_{00})_{j}] 
\quad,
\label{1}
\eeq
where $c_{00}$ and $d_{00}$ are the time-like components of the CPT-even 
flavour-dependent parameters of the fermion sector of the 
SM Lorentz-violating extension \cite{Colladay2}

\beq
{\cal L}^{\rm CPT-even}_{\rm Fermion} = \half i c_{\mu\nu} \overline{\psi}
\gamma^{\mu} \rl \psi + \half i d_{\mu\nu} \overline{\psi} \gamma_5
\gamma^{\mu} \rl  \psi 
\quad,
\label{2}
\eeq
the $\pm$ signs in Eq. (1) indicate the fact that parameter 
$ d_{\mu\nu}$ depends on the chirality of the particles, and $D$ is 
the proper distance of a faraway source which is given in 
terms of its redshift, $z$, Hubble's constant, $H_0$ and the deceleration 
parameter, $q_0$, by the well known expression:

\beq
D = {c \over H_0} [z - {1 \over 2} (1 + q_0) z^2 + O(z^3)] 
\quad.
\label{2a}
\eeq

Time delay (1) arises from the modifications in the dispersion 
relation introduced by the Lorentz violating parameters 
and from the fact they imply in a 
maximum attainable velocity for each particle, namely 
$c_i = c [1 - (c_{00} \pm d_{00})_{i}]$, for massless particles 
\footnote{There are strong reasons not to alter the gauge sector 
of the SM, therefore the electromagnetic radiation is assumed still 
to propagate at the velocity of light (see \cite{Bertolami1} 
and references therein).} or in the 
limit $m << p, E$. Hence, an astrophysical test of Lorentz invariance would 
involve the measurement of the time delay of signals emitted 
by farway sources. 
Particularly suitable candidates are $TeV$ gamma-ray 
flares and the ensued emission of neutrinos by AGN or 
GRBs. It is remarkable, in this respect, that large area ($\sim km^2$) 
high-energy neutrino telescopes are under construction and will possibly 
allow for obtaining information 
that is congenitally correlated with the gamma radiation emitted by 
AGN and GRBs. 

A possible violation of Lorentz invariance may resurrect alternative theories 
of gravity that have genetically built in this feature. Given that the 
post-Newtonian parametrization is the most effective way to compare the 
effects of different theories of gravity with the observational data one 
should consider the alternative theory of gravity that most closely 
resembles General Relativity, the theory that fits most accurately all 
known data. Having this criteria in mind the most suitable alternative model
for gravity that presents preferred-frame effects is Rosen's bimetric 
theory \footnote{From a more theoretical point of view, one could argue 
that some unification schemes such as, for instance string theory, 
are bimetric 
theories.} \cite{Rosen}. Indeed, this theory shares with General 
Relativity the same values for all post-Newtonian parameters \cite{Will}

\beq
\beta = \gamma = 1~~;~~\alpha_{1} = \alpha_{3} 
= \zeta_{1} =  \zeta_{2} = \zeta_{3} = \zeta_{4} = \xi = 0
\quad,
\label{3}
\eeq   
except for parameter $\alpha_{2}$ that signals the presence preferred-frame 
effects (Lorentz violation) in $g_{00}$ and $g_{0i}$ components of the metric. 
Naturally, this parameter vanishes in General Relativity, but in 
Rosen's bimetric theory it assumes the value

\beq
\alpha_{2} = {f_0 \over f_1} - 1
\quad,
\label{4}
\eeq   
where $f_0$ and $f_1$ are the asymptotic values of the components of the 
metric in the Universe rest frame, i.e. $g_{\mu \nu}^{(0)} = 
diag(- f_0, f_1, f_1, f_1)$, 
that is presumably close to the Minkowski metric, although not 
necessarily the same. 
A non-vanishing $\alpha_{2}$ (and $\alpha_{1}$) also implies that angular 
momentum is not conserved and hence that rotational symmetry is lost. 
Bounds on this parameter can be obtained from the 
implied anomalous torques on the Sun, whose absence reveal that 
$\alpha_{2} < 4 \times 10^{-7}$ \cite{Nordtvedt} \footnote{The other 
parameters that imply in preferred-frame effects are bound by pulsar 
PSR J2317+1439 data, that yield, $\alpha_{1} < 2 \times 10^{-4}$ 
\cite{Will}, and 
from the average on the pulse period of millisecond  
pulsars, that gives origin to the impressive bound, 
$\alpha_{3} < 2.2 \times 10^{-20}$ \cite{BellD}.}.
 
It is worth recalling that the most conspicuous feature of Rosen's 
theory is that 
it contains, besides the metric tensor field, a nondynamical metric, 
$\eta_{\mu \nu}$, which in 
typical theories is chosen to be Riemann flat everywhere in spacetime.
The relevant field equations can be obtained from the action:

\beq
S = {1 \over 64 \pi G} \int d^4x \sqrt{-\eta}~\eta^{\mu \nu} g^{\alpha \beta} 
g^{\gamma \delta} \left[g_{\alpha \gamma \vert \mu} 
g_{\beta \delta \vert \nu}  
- {1 \over 2} g_{\alpha \beta \vert \mu} g_{\gamma \delta \vert \nu}\right] 
+ S_{Matter}(\Phi, g_{\mu \nu})
\quad,
\label{5}
\eeq   
where the vertical line ``$\vert$'' 
denotes covariant derivatives with respect to 
$\eta_{\mu \nu}$ and $\Phi$ a generic matter field. The relationship 
between Newton's constant and the coupling constant in action (6) is
$G_{N} = G (f_0 f_1)^{1/2}$.

Another relevant fact in Rosen's bimetric theory is that, in vacuum, 
the linearized wave equation in the weak field limit, 
$\vert h_{\mu \nu}\vert << \vert g_{\mu \nu}\vert$, reads 

\beq
{f_0 \over f_1}~h_{\mu \nu, 00} - c^{2} \nabla ^{2} h_{\mu \nu} = 0
\quad,
\label{6}
\eeq 
implying that gravitational waves propagate in Rosen's theory with velocity
$c(f_1/f_0)^{1/2}$. 

Thus, given that AGN and GRBs phenomena are believed to be powered 
by massive black holes, a time delay measurement of signals of 
gamma-ray flares and neutrinos 
from AGN or GRBs as described above could 
be further correlated with the arrival of the 
gravitational radiation generated by that phenomena. 
Measurements of this nature could reveal whether 
the post-Newtonian parameter $\alpha_2$ is non-vanishing or, at least, 
may allow improving the present bound. 

We further mention that studies of Rosen's bimetric theory in the context 
of homogeneous and isotropic cosmological models, where all relevant 
quantities and parameters are functions of cosmic time, 
indicate that some flat models do expand from a singular 
state at a finite proper time in the past and imply for the time 
variation of the gravitational 
coupling (see \cite{Will} and references therein)

\beq
\left({\dot{G} \over G} \right)_{0} \gsim 0.51~H_{0}~[1 + 
3~\Omega_{M0} (1 + \alpha_{20})^{-1}]  
\quad,
\label{7}
\eeq
where $\Omega_{M0}$ is the density parameter of matter at present. This 
variation is still
consistent with data \cite{Will} provided that, at present, 
$|\alpha_{20}| << 1$, which is compatible with the abovementioned bound. 
This prediction can be tested by future experiments 
(see e.g. \cite{Will}).

To summarize we could say that the observation of UHECRs with energies beyond 
the GZK cut-off leads us to seriously consider the breaking of Lorentz 
invariance. This breaking may have its origin either in 
non-trivial solutions of string field theory or in the dynamics of a 
$3$-brane model of our world. We have outlined a strategy to perform 
an independent 
astrophysical test of Lorentz symmetry via the measurement of the time 
delay in the arrival of 
signals of gamma radiation and neutrinos emitted by farway sources such as 
AGN and GRBs. We have also suggested that further correlating the gamma-ray 
and neutrino information with the one from the gravitational radiation  
emitted by the same sources may turn out to be helpful in improving 
the existing bound on 
the post-Newtonian parameter $\alpha_2$ and verifying a 
possible breaking of rotational symmetry. 
Finally, we stress that a non-trivial result 
of the astrophysical test we 
have proposed would enable us to have an insightful glimpse 
into the brave new physics beyond the SM.

\vfill
\newpage


\end{document}